\documentclass[runningheads]{llncs}
\usepackage[T1]{fontenc}
\usepackage{graphicx}

\usepackage[caption=false,font=footnotesize,labelfont=sf,textfont=sf]{subfig}
\usepackage{balance}
\usepackage{amssymb}
\usepackage{amsmath,amssymb,amsfonts}
\usepackage{algpseudocode}
\usepackage{algorithm}
\algnewcommand{\algorithmicor}{\textbf{ or }}
\algnewcommand{\algorithmicand}{\textbf{ and }}
\algnewcommand{\OR}{\algorithmicor}
\algnewcommand{\AND}{\algorithmicand}
\usepackage{adjustbox}
\usepackage{booktabs}
\usepackage[binary-units=true]{siunitx}
\usepackage{threeparttable}
\usepackage{enumitem}
\usepackage{subfig}
\usepackage[font=footnotesize]{caption}
\usepackage[utf8]{inputenc}
\newcommand{\proposedapp}{AnTi-MiCS}

\usepackage{orcidlink}
\usepackage{fancyhdr}
\usepackage{xcolor}

\fancypagestyle{firstpage}{
  \fancyhf{} % clear all header and footer fields
  \fancyfoot[L]{\textcolor{blue}{%
    \parbox{0.99\textwidth}{\footnotesize
    This paper is accepted to be published in International Conference on Embedded Computer Systems: Architectures, Modeling and Simulation (SAMOS XXV) 2025.
    }%
  }}

}

\begin{document}

% \onecolumn
% \thispagestyle{empty}
% \input{DACReview.tex}

% \twocolumn
% \setcounter{page}{1}
% \setcounter{figure}{0}

%\titlerunning{Analytical Framework for Bounding Time in Embedded MC Systems}
\title{AnTi-MiCS: \underline{An}alytical Framework for Bounding \underline{Ti}me in Embedded \underline{Mi}xed-\underline{C}riticality \underline{S}ystems }

\author{Behnaz Ranjbar\orcidlink{0000-0001-7944-7101} 
\and
Akash Kumar\orcidlink{0000-0001-7125-1737}
}
\authorrunning{B. Ranjbar and A. Kumar}
\institute{Chair of Embedded Systems, Ruhr University Bochum, Germany\\
\email{\{behnaz.ranjbar,akash.kumar\}@rub.de}\\
%\url{http://www.springer.com/gp/computer-science/lncs} \and ABC Institute, Rupert-Karls-University Heidelberg, Heidelberg, Germany\\ \email{\{abc,lncs\}@uni-heidelberg.de}
}

\maketitle

% \maketitle
\thispagestyle{firstpage}

\begin{abstract}
    In Mixed-Criticality~(MC) systems, although the high Worst-Case Execution Time~(WCET) serves as a conservative upper bound representing the task's maximum execution time under all conditions, obtaining a low WCET is essential for representing realistic executions and improving utilization and Quality-of-Service~(QoS).
Nevertheless, determining appropriate low WCET(s) for lower-criticality (LO) modes poses a significant challenge. 
Opting for a very low value of this WCET enhances processor utilization by scheduling more tasks in LO mode. Conversely, employing a larger WCET ensures fewer mode switches, thereby enhancing QoS, albeit at the cost of processor utilization. 
This paper proposes an analytical approach, AnTi-MiCS, to determine the appropriate low WCET through design-time analysis of task executions. In some cases, a single low WCET may not be adequate to capture large variations in the execution time distribution, for example, in scenarios like bimodal distributions. 
Therefore, we further propose a scalable approach, MulTi-MiCS, to compute multiple appropriate low WCETs. This approach exploits the temporal correlation between subsequent inputs presented to the application. 
Experimental results, conducted on a real platform with embedded real-time benchmarks, demonstrate the efficacy of our proposed scheme, in which QoS is improved by 30.27\% on average while reducing utilization waste by 35.89\%, compared to existing approaches. Besides, MulTi-MiCS improves QoS by 6.41\% compared to AnTi-MiCS while reducing utilization waste by 8.23\%.

\keywords{
Mixed-Criticality, Mode Switching Probability, Quality-of-Service, Timing Analysis, Worst-Case Execution Time~(WCET)
}
\end{abstract}
% \end{IEEEkeywords}

% \maketitle

% \IEEEpeerreviewmaketitle

\section{Introduction}
\label{Sec:Intro}
\vspace{-6pt}
Mixed-Criticality (MC) systems integrate numerous real-time tasks of varying criticality levels on a shared hardware platform to meet stringent requirements concerning cost, space, and timing~\cite{baruah12,burns2022mixed,Ranjbar21,Liu2018}. Safety-critical applications, including automotive and avionics, have transitioned towards these MC~systems in which ensuring correct execution of tasks categorized as Higher-Criticality~(HC) under all conditions is essential to avert catastrophic outcomes, while the execution of more Low-Criticality~(LC) tasks is essential to enhance Quality-of-Service~(QoS), thus improving processor utilization~\cite{burns2022mixed,Ranjbar21,Ranjbar2023dac}.

In MC systems, multiple Worst-Case Execution Times (WCETs) are ascertained, corresponding to the various criticality levels~\cite{baruah12,burns2022mixed,Ranjbar21,Liu2018}. If there are two criticality levels for tasks, two WCETs (named $\textit{WCET}^{\textit{LO}}$ and $\textit{WCET}^{\textit{HI}}$) are determined for each task. $\textit{WCET}^{\textit{HI}}$ represents a conservative boundary, denoting the maximum execution time of a task under all circumstances. \cite{Wilhelm08} summarized many tools and approaches to determine this pessimistic WCET. However, this upper bound is often excessive, and its utilization for task scheduling can result in processor under-utilization and limiting the number of schedulable LC tasks~\cite{Liu2018,Ranjbar21}. Therefore, $\textit{WCET}^{\textit{LO}}$~($\leq$ $\textit{WCET}^{\textit{HI}}$) is determined %, which should closely align with the execution time of the majority of task instances 
to enhance utilization and QoS. At run-time, the system commences operation in low-criticality mode (LO mode). If the execution time of at least one HC task exceeds its $\textit{WCET}^{\textit{LO}}$, the system transitions to the high-criticality mode (HI mode). In this mode, HC tasks are scheduled using $\textit{WCET}^{\textit{HI}}$ to meet deadlines, while LC tasks are either dropped or degraded~\cite{burns2022mixed,Ranjbar21,Liu2018}.
%to ensure the proper execution of HC tasks within their deadlines, the system considers $\textit{WCET}^{\textit{HI}}$ for scheduling HC tasks, and consequently, LC tasks are either dropped or degraded~\cite{burns2022mixed,Ranjbar21,Liu2018}.% to their minimum service requirements~\cite{burns2022mixed,Liu2018,Ranjbar21}.

A large gap between $\textit{WCET}^{\textit{LO}}$ and $\textit{WCET}^{\textit{HI}}$ allows more tasks to be scheduled at design-time; however, may lead to frequent mode switches and dropping more LC tasks at run-time. Conversely, a smaller gap reduces mode switches but limits processor utilization due to scheduling fewer tasks at design-time~\cite{Ranjbar21,Ranjbar2023dac,Ranjbar2021-tcad}. Thus, analyzing and determining an appropriate value of $\textit{WCET}^{\textit{LO}}$ for each task is crucial and challenging for efficient MC system design, one of the aspects that we address in this article. If $\textit{WCET}^{\textit{LO}}$ are not close to the execution times of the majority of task samples, it can lead to significant performance loss for LC tasks or result in processor under-utilization. In general, these task's execution times depend on task's input values. Due to spatial or temporal correlations inherent in the input data stream, as observed in scenarios such as video processing, the execution times of tasks frequently exhibit temporal correlations. As a result, the other main issue in designing an efficient MC system, mentioned in~\cite{burns2022mixed}, is how many $\textit{WCET}^{\textit{LO}}$ levels are required/useful for the same application.

\begin{figure}[t]
	\centering
    \vspace{-5pt}
	\begin{minipage}[t]{1\linewidth}
		\centering
		\includegraphics[width=0.7\columnwidth]{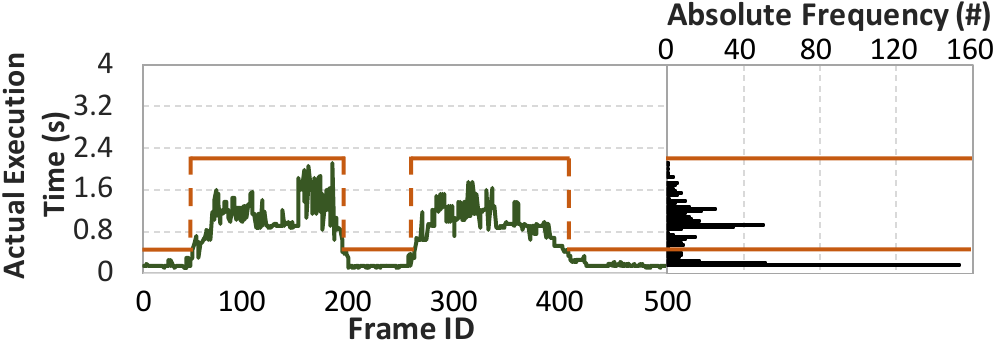}	
        \vspace{-6pt}
		\caption{Execution times for time-recorded videos used as input to the Object Detection function, along with their distribution, highlight the need to consider multiple timing bounds. This underscores the importance of accounting for run-time input behavior and temporal correlation when designing MC systems and determining task properties.}
        %Execution time values for time recording videos used as input for the Object Detection function during run-time, alongside their time distribution. It demonstrates the necessity of considering multiple bounding times, taking into account the run-time behavior of inputs and their temporal correlation for the design of MC systems and the determination of task properties.
		\label{fig:MotivExp}
	\end{minipage}
    \vspace{-15pt}
\end{figure}

\textbf{\textit{Motivational Example:}} To motivate the research problem, the Object Detection function, a key component in autonomous driving, is executed on an ODROID-XU4 board. %This function receives inputs in the form of motion JPEGs converted from videos captured by a road camera during different time slots, experiencing both heavy and light traffic. 
It processes motion JPEGs converted from road camera videos captured during various time slots, covering both heavy and light traffic conditions.
Fig.~\ref{fig:MotivExp} depicts the computational times associated with executing this function, alongside their time distribution. 
As can be seen, the computation times for object detection exhibit variation during run-time and are clustered owing to the temporal correlation between successive inputs provided to the application. By considering one static $\textit{WCET}^{\textit{LO}}$, for example, 2.18\textit{s} in this example, followed by the related works~\cite{baruah12,Liu2018,Ranjbar21}, the system performs well for some time period. However, it may cause under-utilization when a few objects need to be detected, like light traffic, or frequent mode switches during periods with many objects to detect, such as during heavy traffic. Consequently, considering multiple levels of $\textit{WCET}^{\textit{LO}}$ for each task based on the functional behavior and input's temporal correlation is needed to significantly improve the QoS and utilization. Although there is a run-time approach~\cite{Ranjbar2023dac}, that sets $\textit{WCET}^{\textit{LO}}$ at run-time based on the generated dynamic slack and system behavior, this approach confronts two main issues: 1)~The amount of utilization and QoS improvement is uncertain at design-time, while it is critical to ascertain this information for some MC systems, 2)~timing overhead of the run-time approach to set the new $\textit{WCET}^{\textit{LO}}$.
Thus, it is essential to analyze application behavior and its time distribution at design-time, adjusting $\textit{WCET}^{\textit{LO}}$ levels based on the variations in the execution time distribution and input temporal correlation, for instance, $\textit{WCET}^{\textit{LO}}$=0.38\textit{s} in light traffic and $\textit{WCET}^{\textit{LO}}$=2.18\textit{s} in heavy~traffic. 

This paper proposes a framework that introduces a novel analytical approach to evaluate the application time distributions and defines a new metric to determine the appropriate value of $\textit{WCET}^{\textit{LO}}$ for HC tasks to improve utilization and QoS while reducing mode switches. 
Nevertheless, contingent upon the application's functionality, there may indeed exist large variations in the execution time distribution and potential temporal correlations among consecutive inputs presented to the application. Consequently, it becomes apparent that using a single low WCET for each application may not effectively adapt to application input behavior changes, thus limiting the potential for improving QoS. Therefore, we then propose a distribution-based analytical model to obtain diverse $\textit{WCET}^{\textit{LO}}$ to improve utilization and QoS at design-time. To our knowledge, this is the first work that determines~(diverse)~$\textit{WCET}^{\textit{LO}}$ based on application input behavior and time distribution variations while ensuring real-timeliness, improving QoS, and reducing mode switches without incurring any timing overhead at run-time. %In addition, the other key benefit of the proposed method is its applicability to systems with more than two criticality levels, allowing for the determination of distinct WCET levels to be utilized in each criticality mode.

\textbf{\textit{Contributions:}} The main contributions of this paper include:

\begin{itemize}[leftmargin=*]
    \item Proposing an analytical approach, named \textit{AnTi-MiCS}, to analyze the MC applications and obtain an appropriate $\textit{WCET}^{\textit{LO}}$ by defining a new metric based on the execution time distribution.
    \item Presenting a distribution-based scalable approach, named \textit{MulTi-MiCS}, to obtain diverse $\textit{WCET}^{\textit{LO}}$s based on the application’s functionality, input behavior, and time distribution.
    \item Enhancing utilization and QoS by leveraging execution time distribution without incurring online timing overhead, thereby reducing mode switches and ensuring minimum QoS in HI mode.
    % \item  ...
\end{itemize}

The paper is structured as follows: related works and the system model are discussed in Sections~\ref{Sec:RelWork} and~\ref{Sec:Model}, respectively. In Section~\ref{Sec:PPMethod}, we provide the proposed method. Then, we evaluate experiments in Section~\ref{Sec:Evaluation}, and conclude in Section~\ref{Sec:Conclude}.

\vspace{-2pt}
\section{Related Work}
\label{Sec:RelWork}
\vspace{-6pt}
Burns and Davis~\cite{burns2022mixed} recently conducted a comprehensive study of related works in the field of MC system design and identified several open challenges, including the two issues discussed in Section~\ref{Sec:Intro}. To address them, we focus on papers, targeting WCET adjustment and QoS improvement, summarized in Table~\ref{tab:realtedworks}.

\begin{table*}[t]
    \vspace{-10pt}
    \caption{State-of-the-art MC approaches overview}
    \footnotesize
	\centering
	% \vspace{-5pt}
    \adjustbox{max width=\linewidth}{
    \begin{tabular}{cccccc}
    \hline
        %\toprule
        \multicolumn{1}{c}{\centering $\#$} &
		\multicolumn{1}{c}{\centering Related Works} &
		\multicolumn{1}{c}{\centering QoS } &
        \multicolumn{1}{c}{\centering $\textit{WCET}^{\textit{LO}}$} &
		\multicolumn{1}{c}{\centering Scalable} &
		\multicolumn{1}{c}{\centering No Online } %&
		% \multicolumn{1}{c}{\centering Use of}
  \\

          %\toprule
        \multicolumn{1}{c}{\centering } &
		\multicolumn{1}{c}{\centering } &
		\multicolumn{1}{c}{\centering Improv.} &
        \multicolumn{1}{c}{\centering Adjust.} &
		\multicolumn{1}{c}{\centering $\textit{WCET}^{\textit{LO}}$} &
		\multicolumn{1}{c}{\centering Timing Overhead} %&
		% \multicolumn{1}{c}{\centering Use of}
  \\
		
		% \multicolumn{1}{c}{\centering } &
		% \multicolumn{1}{c}{\centering } &
		% \multicolumn{1}{c}{\centering Improv.} &
  %       \multicolumn{1}{c}{\centering Adjust.} &
		% \multicolumn{1}{c}{\centering $WCET^{LO}$} &
		% \multicolumn{1}{c}{\centering Timing Overhead}% &
		% % \multicolumn{1}{c}{\centering ML}
  % \\
		
        \hline
        \hline
        
        \multicolumn{1}{c}{\centering 1} &
        \multicolumn{1}{c}{\centering Baruah'12~\cite{baruah12}, Liu'18~\cite{Liu2018}, Liu'16~\cite{Liu2016}} &
		\multicolumn{1}{c}{\centering $\times$} &
        \multicolumn{1}{c}{\centering $\times$} &
		\multicolumn{1}{c}{\centering $\times$} &
		\multicolumn{1}{c}{\centering $\checkmark$} %&
		% \multicolumn{1}{c}{\centering $\times$}
  \\

        \hline

        \multicolumn{1}{c}{\centering 2} & 
		\multicolumn{1}{c}{\centering 
        % Ranjbar'21a~\cite{Ranjbar21}, 
        Ranjbar'21~\cite{Ranjbar21}, Ranjbar'22a~\cite{Ranjbar2021-tcad}, Ranjbar'24~\cite{Ranjbar2024}, Kumar'24~\cite{Kumar2024}} &
		\multicolumn{1}{c}{\centering $\checkmark$} &
        \multicolumn{1}{c}{\centering $\checkmark$} &
		\multicolumn{1}{c}{\centering $\times$} &
		\multicolumn{1}{c}{\centering $\checkmark$} %&
		% \multicolumn{1}{c}{\centering $\times$}
  \\
         
        \hline
        
        \multicolumn{1}{c}{\centering 3} & 
		\multicolumn{1}{c}{\centering Guo'15~\cite{Guo2015}, Baruah'16~\cite{Baruah2016}, 
        Guo'18~\cite{Guo2018}, Yang'19~\cite{Yang2019} } &
		\multicolumn{1}{c}{\centering $\checkmark$} &
        \multicolumn{1}{c}{\centering $\times$} &
		\multicolumn{1}{c}{\centering $\times$} &
		\multicolumn{1}{c}{\centering $\checkmark$} %&
		% \multicolumn{1}{c}{\centering $\times$}
  \\

        \hline
        
        \multicolumn{1}{c}{\centering 4} & 
		\multicolumn{1}{c}{\centering Gu'16~\cite{Gu2016}, Gu'18~\cite{Gu2018}, Hu'19~\cite{hu2019ffob} } &
		\multicolumn{1}{c}{\centering $\times$} &
        \multicolumn{1}{c}{\centering $\checkmark$} &
		\multicolumn{1}{c}{\centering $\times$} &
		\multicolumn{1}{c}{\centering $\checkmark$} %&
		% \multicolumn{1}{c}{\centering $\times$}
  \\
        
        \hline

        \multicolumn{1}{c}{\centering 5} &
		\multicolumn{1}{c}{\centering Ranjbar'23~\cite{Ranjbar2023dac}} &
		\multicolumn{1}{c}{\centering $\checkmark$} &
        \multicolumn{1}{c}{\centering $\checkmark$} &
		\multicolumn{1}{c}{\centering $\times$} &
		\multicolumn{1}{c}{\centering $\times$} %&
		% \multicolumn{1}{c}{\centering $\checkmark$}
  \\
		
		\hline

  %       \multicolumn{1}{c}{\centering 5} &
		% \multicolumn{1}{c}{\centering Kumar'24~\cite{Kumar2024}} &
		% \multicolumn{1}{c}{\centering $\checkmark$} &
  %       \multicolumn{1}{c}{\centering $\checkmark$} &
		% \multicolumn{1}{c}{\centering $\times$} &
		% \multicolumn{1}{c}{\centering $\checkmark$} %&
		% % \multicolumn{1}{c}{\centering $\checkmark$}
  % \\
		
		% \hline
	
% 		\multicolumn{1}{c}{\centering \makecell{Hu'19~\cite{hu2019ffob}, Ittershagen'15~\cite{Philipp2015}}} &
% 		\multicolumn{1}{c}{\centering $\times$} &
% 		\multicolumn{1}{c}{\centering $\checkmark$} &
% 		\multicolumn{1}{c}{\centering $\times$/$\checkmark$} &
% 		\multicolumn{1}{c}{\centering $\times$}\\
		
% 		\hline
		
		\multicolumn{1}{c}{\centering 6} &
		\multicolumn{1}{c}{\centering Su'16~\cite{Su2016}, Ranjbar'22b~\cite{Ranjbar2022MDPI}
        } &
		\multicolumn{1}{c}{\centering $\checkmark$} &
        \multicolumn{1}{c}{\centering $\times$} &
		\multicolumn{1}{c}{\centering $\times$} &
		\multicolumn{1}{c}{\centering $\times$} %&
		% \multicolumn{1}{c}{\centering $\times$}
  \\
		
		\hline

        \multicolumn{1}{c}{\centering 7} & 
		\multicolumn{1}{c}{\centering Proposed Approach } &
		\multicolumn{1}{c}{\centering $\checkmark$} &
		\multicolumn{1}{c}{\centering $\checkmark$} &
        \multicolumn{1}{c}{\centering $\checkmark$} &
		\multicolumn{1}{c}{\centering $\checkmark$}% &
		% \multicolumn{1}{c}{\centering $\checkmark$}
  \\
		
		\hline
         
    \end{tabular}
    }
    \label{tab:realtedworks}
  \vspace{-10pt}
\end{table*}

From the perspective of $\textit{WCET}^{\textit{LO}}$ determination, most papers like~\cite{baruah12,Liu2018} (row~1) set $\textit{WCET}^{\textit{LO}}$ as a percentage of $\textit{WCET}^{\textit{HI}}$. This policy is inaccurate, leading to wasted system utilization or frequent mode switches, which can disrupt LC task execution. Thus, researchers in~\cite{Ranjbar21} and its extended version~\cite{Ranjbar2021-tcad}~(row~2) have proposed a theoretical scheme to obtain $\textit{WCET}^{\textit{LO}}$ based on Chebyshev's theorem by using average execution times and standard deviation of samples. However, this scheme determines $\textit{WCET}^{\textit{LO}}$ and the probability of overrunning it pessimistically, which can result in under-utilization at run-time. Moreover, it overlooks possible correlations among application inputs. Besides, researchers in~\cite{Ranjbar2024,Kumar2024} have presented Graph-Neural-Network (GNN)-based approaches to determine $\textit{WCET}^{\textit{LO}}$ for independent/dependent MC tasks to improve QoS and reduce mode switching probability; however, these GNN-based approaches 1) may obtain inaccurate $\textit{WCET}^{\textit{LO}}$ due to relying on the use of learning from data, which can be incomplete or biased (i.e., poor training data), 
2) lack of interpretability makes it difficult to understand the rationale behind specific $\textit{WCET}^{\textit{LO}}$ estimates.
%2) can be challenging to interpret, making understanding the reasoning behind a particular $\textit{WCET}^{\textit{LO}}$ estimate difficult. 
These limitations make the GNN-based approach less reliable for MC applications where safety and predictability are paramount. 
In row~3, researchers in~\cite{Guo2015} have presented an Earliest Deadline First (EDF)-based scheduling algorithm using a new task parameter-the probability of exceeding $\textit{WCET}^{\textit{LO}}$ within an hour of execution- to improve schedulability, they do not discuss in detail how this parameter is calculated. Furthermore, the determination is not utilized to enhance the behavior of MC systems. Besides, some works like~\cite{Baruah2016}, propose MC scheduling algorithms to improve schedulability by increasing the utilization that can be assigned to LC and HC tasks and improving QoS; however, they do not address $\textit{WCET}^{\textit{LO}}$ determination or mode switching probability. Additionally, to improve QoS and reduce LC task degradation in HI mode, different scheduling policies have been presented~\cite{Guo2018,Yang2019} to ensure the execution of the defined minimum number of LC tasks instances in HI mode. However, there is still no discussion on determination of $\textit{WCET}^{\textit{LO}}$ and improving the probability of mode switching.

%In addition, s
Some approaches~\cite{Gu2016,Gu2018,hu2019ffob}~(row~4) determined $\textit{WCET}^{\textit{LO}}$ at run-time based on the application behavior. However, these approaches postpone the mode switches for a long time while ensuring the minimum QoS. Indeed, they do not improve QoS by scheduling more LC tasks. 
Furthermore, researchers in~\cite{Ranjbar2023dac}~(row~5) employ Machine-Learning (ML) techniques to determine $\textit{WCET}^{\textit{LO}}$. However, the ML-based approach in~\cite{Ranjbar2023dac} is a run-time approach with associated timing overhead, $\textit{WCET}^{\textit{LO}}$s are determined only if sufficient dynamic slack exists on the processor, and thus the amount of QoS improvement is also unknown at design-time. %Although \cite{Kumar2024} have presented design-time approach, ..
% In contrast, our scheme concentrates on deriving $WCET^{LO}$ based on the timing distribution behavior of tasks at design-time, independent of whether dynamic slack would be generated at run-time. 
In addition,~\cite{Su2016,Ranjbar2022MDPI}~(row~6) have adopted a similar policy of row~1 in determining $\textit{WCET}^{\textit{LO}}$ to enhance QoS at run-time, leveraging the dynamic slack resulting from early completion of HC tasks; however, although these approaches have online timing overhead, as dynamic slack is treated as a wrapper task with a deadline~\cite{Su2016}, it cannot be utilized freely at any time. Consequently, these approaches do not optimally utilize system idle times to enhance QoS. Also, there is no estimate of QoS improvement at design-time.

To this end, an analytical approach is needed to determine the useful diverse $\textit{WCET}^{\textit{LO}}$s and confidence in them to identify the amount of utilization and QoS improvement at design-time while overcoming run-time timing overhead. 
We propose a novel framework that initiates an analysis of the application's timing distribution at design-time and determines appropriate scalable $\textit{WCET}^{\textit{LO}}$s while enhancing QoS and reducing the mode switching probability.

% graceful degradation in MCS
% \cite{Guo2018},\cite{Yang2019} like fantom
% \cite{Baruah2016} like \cite{Guo2015} improving schedulability

\vspace{-5pt}

% same formula Eq. 1? \cite{Guo2015}

% ML-based WCET estimation \cite{Ranjbar2024,Kumar2024}

\vspace{-2pt}
\section{Mixed-Criticality System Model}
\label{Sec:Model}
\vspace{-6pt}
In this paper, real-time MC applications are considered that comprise periodic independent tasks and are executed on a preemptive uni-processor. Similar to the approach outlined in~\cite{baruah12,Liu2018,Ranjbar21}, we examine a dual-criticality system where a set of MC tasks~$\{\tau_1, \tau_2, ..., \tau_n\}$ are considered, and each task $\tau_i$ is characterized as~$\{\zeta_i, \textit{WCET}^{\textit{LO}}_i,$ $\textit{WCET}^{\textit{HI}}_i, D_i, P_i\}$. $\zeta_i\in \{LC,HC\}$ denotes the criticality level of the task. $\textit{WCET}^{\textit{HI(LO)}}_i$ denotes the WCETs of the task in the HI~(LO) mode. $D_i$ and $P_i$ denote the deadline and period, respectively, where analogous to many MC works such as~\cite{baruah12,Liu2018,Ranjbar21}, in this paper $D_i=P_i$.

From the perspective of MC system operational model, execution starts in LO mode, where both HC and LC tasks are guaranteed to meet their deadlines. If the execution time of any HC task exceeds its $\textit{WCET}^{\textit{LO}}$, the system switches to HI mode, where LC tasks are dropped or degraded to minimum service levels to guarantee HC task deadlines. If there are no ready HC tasks in the processor's queue, the system safely reverts back to~the~LO~mode~\cite{baruah12,Liu2018,Ranjbar21}. Besides,
from the LC tasks’ QoS perspective, as a general rule, the system should schedule and execute the LC tasks, considering their actual periods, to enhance QoS and ensure high output precision. However, when the system switches to the HI mode, the LC tasks and consequently, the QoS are degraded by regulating the execution rate for LC tasks, specifically by adjusting the LC tasks' periods~\cite{Su2016}. If we define QoS as $\frac{\#Schd-LC}{\#all-LC}$~\cite{Ranjbar2023dac,Ranjbar2021tcad} where $\#\textit{all-LC}$ and $\#\textit{Schd-LC}$ are the numbers of all LC tasks and scheduled LC tasks, respectively, scheduling all LC tasks corresponds to a state where all LC tasks can be consistently released according to their actual periods. Consequently, $\#\textit{Schd-LC}$ can be implemented by scheduling the release of LC tasks with an extended period.

\vspace{-2pt}
\section{Proposed Method}
\label{Sec:PPMethod}
\vspace{-6pt}
% As previously stated, determining the appropriate $WCET^{LO}$ for HC tasks poses a significant challenge during the design of MC systems. 

To address the challenge of determining the appropriate $\textit{WCET}^{\textit{LO}}$ for HC tasks, enhance schedulability, and enable the execution of more LC tasks, we first propose \textit{AnTi-MiCS (\underline{An}alytical bounding \underline{Ti}me for \underline{Mi}xed-\underline{C}riticality \underline{S}ystems}), introducing a metric to drive $\textit{WCET}^{\textit{LO}}$ from the application's time distribution. 

%In order to address the challenge of determining the appropriate $WCET^{LO}$ for HC tasks and enhance task schedulability and enable the execution of more LC tasks, the proposed scheme designs the MC systems by analyzing the MC applications at design-time and chooses diverse appropriate WCETs for each HC task, depending on their input behavior and functionalities. This strategy enhances task schedulability and enables the execution of more LC tasks, leveraging the behavior of application inputs (e.g., temporal correlation) and the significant disparity between actual execution times and WCET. To determine $WCET^{LO}$s, we first propose our analytical time bounding approach, named \textit{AnTi-MiCS: \underline{An}alytical bounding \underline{Ti}me for \underline{Mi}xed-\underline{C}riticality \underline{S}ystems}, by introducing a metric based on the application time distribution.

\vspace{-2pt}
\subsection{AnTi-MiCS in Detail}
\label{subsec:PPMethod1}
\vspace{-4pt}

%%Figure~\ref{fig:overview} shows an overview of proposed framework, in which each benchmark, $WCET^{HI}$ is obtained by employing OTAWA~\cite{OTAWA2010}, minimum number of application executions on the ODROID board is determined using the method outlined in~\cite{Ranjbar2021-tcad}, the time distribution of application is employed by running the application on ODROID board. $WCET^{LO}$ and percentage of overrunning it~($P^{ovrun}$) are obtained by using proposed approach, through using Eq~(\ref{eq:Metric1}), which equals to Eq.~(\ref{eq:multimetric}) when no $WCET^{LO}$ level is determined. Now, to determine more $WCET^{LO}$ levels, the obtained parameter is given as input to the proposed approach (i.e., using Eq.~(\ref{eq:multimetric})) to obtain the next level of $WCET^{LO}$.

\begin{figure*}[t]
	    \centering
        \vspace{-10pt}
		\includegraphics[width=0.9\linewidth]{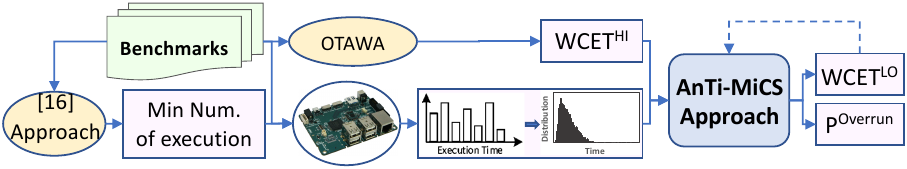}
        \vspace{-7pt}
		\caption{An overview of the proposed approach.}
		\label{fig:overview}
  \vspace{-12pt}
\end{figure*}

Since we target embedded systems, normally, a designer knows, at design time, which applications are expected to execute in embedded systems~\cite{Ranjbar2021-tcad}. Given this, we introduce an analytical approach that leans on an extensive experiment-driven method and is employed at design-time. Fig.~\ref{fig:overview} shows an overview of the proposed framework, in which we obtain $\textit{WCET}^{\textit{HI}}$ of each benchmark by employing OTAWA~\cite{OTAWA2010}. \textcolor{black}{The proposed framework is based on the probabilistic analysis of tasks—the details of what the probabilistic analysis and probability distribution of real-time tasks are, are provided in~\cite{Cazorla2013,Wilhelm08}.}

First, to extract the application's time distribution, a minimum number of executed samples required to obtain an accurate distribution graph should be determined, obtaining using the method in~\cite{Ranjbar2021-tcad}. The application's time distribution is then obtained by running it on ODROID~XU4, powered by ARM~Cortex~A7. 

% probabilistic real-time systems \cite{Cazorla2013,Cucu2012}

\begin{figure}[t]
\centering
\vspace{-10pt}
\subfloat[Frequency distribution]{\label{MethodFig1a}{\includegraphics[width=0.45\columnwidth]{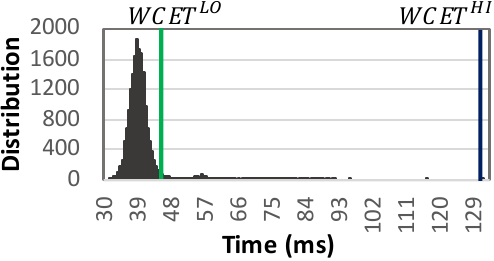}}}
\hfill
\subfloat[\textit{EET} Metric]{\label{MethodFig1b}{\includegraphics[width=0.42\columnwidth]{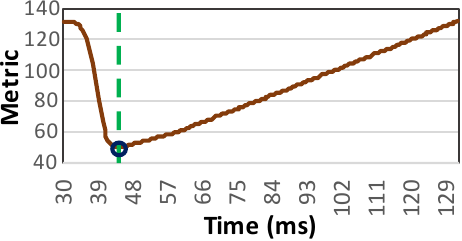}}}
\vspace{-10pt}
\caption{Application analysis which has one peak in its distribution curve.}
\label{MethodFig1}
\vspace{-12pt}
\end{figure}

We clarify how AnTi-MiCS determines the appropriate $\textit{WCET}^{\textit{LO}}$ for each HC task by an example. %in Fig.~\ref{MethodFig1}. 
Fig.~\ref{MethodFig1a} shows execution time distribution~of~the quick-sort application from MiBench~\cite{Guthaus2001}, with 20000 instances running on ODROID XU4. 
%Noted that, the determination of the minimum number of samples required for obtaining an accurate distribution graph can be established using the method outlined in~\cite{Ranjbar2021-tcad}. 
To make a reasonable trade-off between utilization and mode switching probability, the proper $\textit{WCET}^{\textit{LO}}$ should correspond to the minimum time in which the majority of sample execution times are less than it. Therefore, we introduce a metric called \textit{Expected Execution Time (EET)}, presented in Eq.~(\ref{eq:Metric1}).
\begin{equation}
\footnotesize
\label{eq:Metric1}
    EET (t) =\alpha(t)\times t+(1-\alpha(t))\times WCET^{HI} 
\end{equation}
%where
\begin{equation}
\footnotesize
    % \alpha(t)=\frac{\int_{0}^{t}f(t)dt}{\int_{0}^{WCET^{HI}}f(t)dt}
    \alpha(t)=\frac{\#(Samples)|_{ExeTime(Sample)\leq t}}{\#All~Samples}
\end{equation}

In Eq.~(\ref{eq:Metric1}), $\textit{WCET}^{\textit{HI}}$ is constant and obtained using OTAWA~\cite{OTAWA2010}. $\alpha(t)$ represents the ratio of the number of instances with execution times less than \textit{t} to the total number of instances, that their execution times are less than $\textit{WCET}^{\textit{HI}}$. As a result, $\alpha(\textit{WCET}^{\textit{HI}})$=1. %(i.e., the ratio of execution distribution at time \textit{t} to the total execution frequency). 
Eq.~(\ref{eq:Metric1}) represents that \textit{EET} at time \textit{t} equals the scenario in which $\alpha(t)$($\leq1$) of samples complete execution before time \textit{t}, while the remaining (i.e., $\alpha(\textit{WCET}^{\textit{HI}})$-$\alpha(t)$=1-$\alpha(t)$) complete within the range of time \textit{t} and $\textit{WCET}^{\textit{HI}}$. 
Fig.~\ref{MethodFig1b} shows how the \textit{EET} changes by time \textit{t}. Our emphasis is on identifying the minimum time \textit{t}, which encompasses the maximum execution distribution (i.e., the execution times of most instances), thereby improving utilization and reducing mode switches. 
% This approach aims to enhance processor utilization and minimize mode switches. 
Indeed, our goal is to determine the minimum \textit{EET} as it corresponds to the proper $\textit{WCET}^{\textit{LO}}$, presented in Eq.~(\ref{eq:metricdriv}). 
\begin{equation}
\footnotesize
\label{eq:metricdriv}
    WCET^{LO}=t \quad where \quad \frac{dEET(t)}{dt}=0
\end{equation}

To provide a clear demonstration, in Fig.~\ref{MethodFig1}, $\textit{WCET}^{\textit{HI}}=$131\textit{ms}. Considering \textit{t}=44\textit{ms}, $\alpha(t)$=92.81\% and \textit{EET(t)}=50.32. Now, if \textit{t}=55\textit{ms}, $\alpha(t)$=97.1\% and \textit{EET(t)}=57.20. Indeed, in this case, time \textit{t} appears to be more pessimistic and has a higher \textit{EET} value compared to the previous scenario, which leads to less processor utilization for scheduling LC tasks. Avoiding such pessimism is crucial when designing MC systems, as even a 4.29\% increase in $\alpha(t)$ can result in a 25\% increase in execution time. Any values of \textit{t} below 44\textit{ms} result in an increase in mode switches, which is undesirable as it hinders the desired QoS. As a result, obtained $\textit{WCET}^{\textit{LO}}$ in Eq.~(\ref{eq:metricdriv}) represents that for any application, the time corresponding to the minimum \textit{EET} value serves as the proper $\textit{WCET}^{\textit{LO}}$ since it signifies the higher processor utilization that can be assigned to LC tasks, and fewer mode switches~($\textit{P}^{\textit{ovrun}}$=1-$\alpha(\textit{WCET}^{\textit{LO}})$), as observed in experiments. 
Note that although Fig.~\ref{MethodFig1} illustrates a time distribution resembling a normal distribution, we have analyzed various real benchmarks with different time distribution shapes. The proposed approach for determining $\textit{WCET}^{\textit{LO}}$ operates independently of the task's distribution and can be applied to most time distribution types, including normal, skewed, bimodal, and uniform distributions.

\textbf{\textit{Algorithm}:} Algorithm~\ref{alg:PPMethod} outlines the pseudo-code of the AnTi-MiCS. The algorithm takes as inputs the application for which we intend to determine $\textit{WCET}^{\textit{LO}}$, platform, and variable \textit{N}, representing the total number of application samples' executions, established using the method outlined in~\cite{Ranjbar2021-tcad}. The outputs are the application's $\textit{WCET}^{\textit{LO}}$ and the probability of overrunning it~($P^{ovrun}$), which leads to mode switching. The procedure initiates by executing the application \textit{N} times on specified platform to obtain the execution time distribution~(lines~2,3). Then, for each iteration from one to $\textit{WCET}^{\textit{HI}}$, $EET(t)$ is computed~(line~6,7), compared with the variable $EET_{min}$, and if the variable is modified, the time \textit{t} is assigned to $Time^{opt}$~(lines~8,9). This process is executed iteratively until the optimal $Time^{opt}$ is determined and assigned to $\textit{WCET}^{\textit{LO}}$ (line~11). 
As evident in the algorithm, the time complexity in determining $\textit{WCET}^{\textit{LO}}$ for each task is a linear function of the $\textit{WCET}^{\textit{HI}}$ value. As $\textit{WCET}^{\textit{HI}}$ increases, the required time also increases. However, this low timing overhead is negligible, as the proposed method operates at design-time.

\begin{algorithm}[!t]
\footnotesize
\vspace{-2pt}
\caption{\proposedapp~Scheme}
	\label{alg:PPMethod}
    \footnotesize
	{
		\begin{algorithmic}[1]		    
		    \renewcommand{\algorithmicrequire}{\textbf{Input:}}
			\renewcommand{\algorithmicensure}{\textbf{Output:}}					
			\Require Application, N, Platform% Single Processor Platform}
			\Ensure $WCET^{LO}_{App}$, $P^{ovrun}_{App}$
			\Procedure{AnTi-MiCS~Function()}{}      
			    \For {i= 1 to N}
			        $ExeTime_{App}$(i)= $MeasureTime$ (Platform, Application)
                \EndFor
                \State $Dist_{App}$ = $Fuc_{Dist}$ ($ExeTime_{App}$)
                \State $Time^{opt}=WCET^{HI}$, $EET_{min}=\infty$
			    \For {\textit{t}= 1 to $\textit{WCET}^{\textit{HI}}$}	  
			        \State $EET(t) =\alpha(t)\times t+(1-\alpha(t))\times WCET^{HI}$ //Eq.~(\ref{eq:Metric1})
                    \If{$EET_{min}\leq EET(t)$}
                        $EET_{min}= EET(t)$ , $Time^{opt}=t$
                    \EndIf
			    \EndFor
			    \State $WCET^{LO}_{App}=Time^{opt}$, $P^{ovrun}_{App}=1-\alpha(Time^{opt})$
			\EndProcedure
		\end{algorithmic}
	}
\end{algorithm}

\vspace{-2pt}
\subsection{MulTi-MiCS: Scalable AnTi-MiCS}
\vspace{-4pt}
As mentioned in previous sub-section, AnTi-MiCS can be applied to any application with any time distribution to obtain its $\textit{WCET}^{\textit{LO}}$. Observing the time distribution, as illustrated in Fig.~\ref{MethodFig1a}, reveals a single peak in its distribution. However, in the time distribution of some applications, more than one peak might exist, indicating large variations in the execution time distribution, as presented in the motivational example of Section~\ref{Sec:Intro}, leading to multiple local minimums in \textit{EET} metric, i.e., there would be multiple points with $\frac{dEET}{dt}=0$. Consequently, we propose a scalable analytical approach, named MulTi-MiCS, to obtain diverse $\textit{WCET}^{\textit{LO}}$ levels and enhance QoS and utilization by scheduling more LC tasks over some time, while preserving the mode-switching probability. MulTi-MiCS resolves the question of how many $\textit{WCET}^{\textit{LO}}$ levels are required/useful for the same application, as mentioned in Section~\ref{Sec:Intro} and discussed in~\cite{burns2022mixed}. %Note that the efficacy of MulTi-MiCS depends on tasks' inputs behavior and functionalities. 

\begin{figure}[t]
	\centering
    \vspace{-8pt}
	\includegraphics[width=0.72\columnwidth]{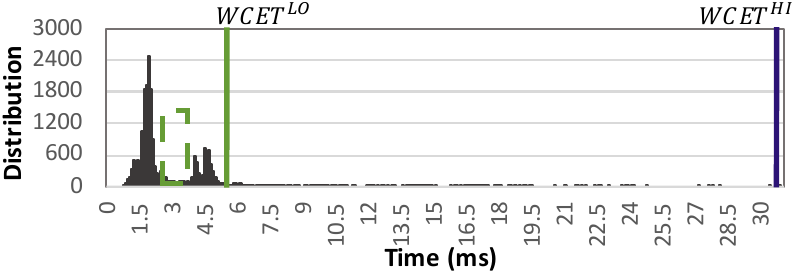}	
    \vspace{-10pt}
	\caption{Application analysis which has more than one peak in its distribution curve.}
	\label{fig:MethodFig2}
 \vspace{-12pt}
\end{figure}

Fig.~\ref{fig:MethodFig2} shows the time distribution of the Epic application from MiBench~\cite{Guthaus2001}, which is run on ODROID-XU4 board. The appropriate $\textit{WCET}^{\textit{LO}}$ is obtained~by AnTi-MiCS, equals 5.2\textit{ms}, with the overrun probability of 4.13\%. 
As can be~seen, the application has more than one peak in its distribution. Depending on the application functionality, each peak may signify temporal correlations between subsequent inputs provided to the application. 
For example, in the case of temporal correlation, in Fig.~\ref{fig:MethodFig2}, one or two more levels of bounding time (which is less than the obtained $\textit{WCET}^{\textit{LO}}$) can be considered. These bounding times can be in a range specified after each peak, shown as a box in this figure. %Considering different bounding times (i.e., diverse $WCET^{LO}$ levels) in designing MC systems can contribute to improve utilization by scheduling more number of LC tasks over a time period. 
As a result, application input behavior and its temporal correlation should be considered when scheduling MC tasks. In the proposed approach, depending on the task's input behavior, different bounding times (i.e., diverse $\textit{WCET}^{\textit{LO}}$ levels) are presented, leading to various QoS improvements by assigning more utilization to LC tasks. 

Now, assume that we have \textit{m} levels of $\textit{WCET}^{\textit{LO}}$~($\textit{WCET}^{\textit{LO,1}}$, $\textit{WCET}^{\textit{LO,2}}$, ..., $\textit{WCET}^{\textit{LO,m}}\}$), where $\textit{WCET}^{\textit{LO,m}}$$\leq...\leq$$\textit{WCET}^{\textit{LO,1}}$ and $\textit{WCET}^{\textit{LO,1}}$ is obtained by AnTi-MiCS. We extend \textit{EET} metric in Eq.~(\ref{eq:Metric1}) to obtain the second level of $\textit{WCET}^{\textit{LO}}$ ($\textit{WCET}^{\textit{LO,2}}$). Eq.~(\ref{eq:2levelmetric}) presents the extended one, called \textit{Scalable Expected Execution Time}~(\textit{SEET}).
{\footnotesize
\begin{align}
\label{eq:2levelmetric}
    SEET(t)=\alpha(t)\times t+ (\alpha(WCET^{LO,1}) -\alpha(t))\times WCET^{LO,1} +  \nonumber \\  
    (1-\alpha(WCET^{LO,1}))\times WCET^{HI}
\end{align}
}
where \textit{t} varies in the range of [0,$\textit{WCET}^{\textit{LO,1}}$], and $0\leq\alpha(t)\leq\alpha(\textit{WCET}^{\textit{LO,1}})\leq1$, and $\alpha(\textit{WCET}^{\textit{LO,1}})$ is obtained by AnTi-MiCS. Following the same argument of Eq.~(\ref{eq:Metric1}), this equation represents that \textit{SEET(t)} is determined by the scenario in which $\alpha(t)$ samples complete execution before time \textit{t} (i.e., [0,t]), $(\alpha(\textit{WCET}^{\textit{LO,1}})$-$\alpha(t))$ of samples finish execution within the range of ($t,\textit{WCET}^{\textit{LO,1}}$], and the remaining samples (i.e., 1-$\alpha(\textit{WCET}^{\textit{LO,1}})$) finish within the range of ($\textit{WCET}^{\textit{LO,1}}$, $\textit{WCET}^{\textit{HI}}$]. Therefore, the second $\textit{WCET}^{\textit{LO}}$ level (i.e., $\textit{WCET}^{\textit{LO,2}}$ which is $\leq\textit{WCET}^{\textit{LO,1}}$) corresponds to the minimum \textit{SEET} value which results in enhanced utilization by scheduling a greater number of tasks, thereby improving the QoS for a timing period. 
For example, for the Epic application, mentioned in Fig.~\ref{fig:MethodFig2}, $\textit{WCET}^{\textit{LO,2}}$=$2.3ms$. By assuming a temporal correlation within inputs, for this example, 64.84\% of the times, $\textit{WCET}^{\textit{LO}}$ can be bounded to 2.3\textit{ms}, and while the input behavior changes and the execution times are more than 2.3\textit{ms}, $\textit{WCET}^{\textit{LO}}$ can equal to $\textit{WCET}^{\textit{LO,1}}$=$5.2$\textit{ms} in 31.203\% of the times. Consequently, when the execution time of a sample exceeds 5.2\textit{ms} ($P^{ovrun}$=$4.13\%$), the system switches to the HI mode. As a result, considering two levels of $\textit{WCET}^{\textit{LO}}$, when $\textit{WCET}^{\textit{LO}}$=$2.3ms$, in 64.84\% of the time, the utilization can be improved by scheduling more LC tasks. It is noteworthy that, given the proximity of the last two peaks, it would be inefficient to designate a separate WCET level between them, a consideration our approach can take into account.

\begin{figure}[t]
	\centering
    \vspace{-9pt}
	\includegraphics[width=0.72\columnwidth]{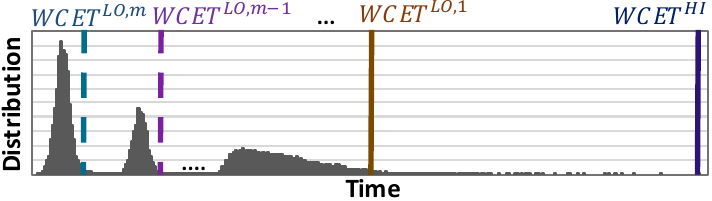}	
    \vspace{-10pt}
	\caption{Application distribution analysis in MulTi-MiCS.}
	\label{fig:multipeak}
 \vspace{-12pt}
\end{figure}

In the presence of multiple peaks and diverse temporal correlations within inputs and needing more than two $\textit{WCET}^{\textit{LO}}$ levels, shown in Fig.~\ref{fig:multipeak}, the same scenario of Eq.~(\ref{eq:2levelmetric}) can be extended to derive more $\textit{WCET}^{\textit{LO}}$ levels, as follows: %depicted in Eq.~(\ref{eq:multimetric}).
% \begingroup\makeatletter\def\f@size{9}\check@mathfonts
% \def\maketag@@@#1{\hbox{\m@th\large\normalfont#1}}
{\footnotesize
\begin{align}
\label{eq:multimetric}
    SEET(t)=\alpha(t)\times t+ \sum_{i=2}^{m}(\alpha(WCET^{LO,i-1})-\alpha(WCET^{LO,i}))  \times WCET^{LO,i-1} \nonumber \\   + (1-\alpha(WCET^{LO,1}))\times WCET^{HI}
\end{align}
}
% \begingroup\makeatletter\def\f@size{10}\check@mathfonts
% \def\maketag@@@#1{\hbox{\m@th\large\normalfont#1}}
where \textit{t} varies in the range [0,$\textit{WCET}^{\textit{LO,m-1}}$] and $\alpha(\textit{WCET}^{\textit{LO,m}})=\alpha(t)$. %In this work, 
This iterative process is continued %until the designer diagnoses to stop based on the task's functionality and its time distribution. In this paper, the process continues
until no further local minimal exists, and the rate of utilization improvement by lowering the $\textit{WCET}^{\textit{LO}}$ level is significant enough (at least 0.05 improvement in utilization ($\textit{WCET}^{\textit{LO,j}}$-$\textit{WCET}^{\textit{LO,j+1}})/period$, where $\textit{WCET}^{\textit{LO,j+1}}$<$\textit{WCET}^{\textit{LO,j}}$), used in our experiment) to enhance QoS.  %In this case that more than $WCET^{LO}$ level is required, the obtained $WCET^{LO}$ level is given as input to the proposed approach (i.e., using Eq.~(\ref{eq:multimetric})) to obtain the next level of $WCET^{LO}$.
% \begin{align}
% \label{eq:multimetric}
%     EET(t)=\alpha(t)\times t+\sum(\alpha(WCET^{LO,l_{i-1}})-\alpha(WCET^{LO,li}))\times \nonumber \\ WCET^{LO,li} + (1-\alpha(WCET^{LO,l0}))\times WCET^{HI}
% \end{align}

%Figure~\ref{fig:overview} shows an overview of analytical approach, in which each benchmark, $WCET^{HI}$ is obtained by employing OTAWA~\cite{OTAWA2010}, minimum number of application executions on the ODROID board is determined using the method outlined in~\cite{Ranjbar2021-tcad}, the time distribution of application is employed by running the application on ODROID board. 
% Now, to determine more $WCET^{LO}$ levels, the obtained parameter is given as input to the proposed approach (i.e., using Eq.~(\ref{eq:multimetric})) to obtain the next level of $WCET^{LO}$.

\textbf{\textit{MC System Operational Model}}: MulTi-MiCS operates in switching between LO and HI modes in a manner analogous to the approaches outlined in~\cite{baruah12,burns2022mixed,Liu2018,Ranjbar21}, as discussed in Section~\ref{Sec:Model}. Indeed, the system switches to the HI mode only if the execution of at least one HC task exceeds its $\textit{WCET}^{\textit{LO}}$, as determined by the AnTi-MiCS approach. (i.e., Eq.~\eqref{eq:Metric1} to~\eqref{eq:metricdriv}), and equals to $\textit{WCET}^{\textit{LO,1}}$, in MulTi-MiCS approach. 
However, our method introduces a distinction by defining multiple bounding times~(i.e., diverse $\textit{WCET}^{\textit{LO}}$s) for each HC task in the LO mode to enhance utilization. Note that the mode switches are not triggered by these diverse $\textit{WCET}^{\textit{LO}}$s~(i.e., $\{\textit{WCET}^{\textit{LO,m}}$, $\textit{WCET}^{\textit{LO,m-1}}$,..., $\textit{WCET}^{\textit{LO,2}}\}$). In the LO mode, depending on a task's input behavior, temporal correlation, different kinds of inputs, and the resulting task execution-times, the system dynamically selects one of the $\textit{WCET}^{\textit{LO}}$ levels for an HC task. In fact, the system opts for the smallest $\textit{WCET}^{\textit{LO}}$ level (obtained at design-time) that is greater than the actual execution-times, while guaranteeing real-timeliness. 

For instance, in Section~\ref{Sec:Intro}, the Object Detection Function's time distribution has two peaks due to the various inputs; leading to two $\textit{WCET}^{\textit{LO}}$ levels, one used for light traffic ($\textit{WCET}^{\textit{LO,2}}$=$0.38s$), and other for heavy traffic ($\textit{WCET}^{\textit{LO,1}}$=$2.18s$). At run-time, suppose the system is in the LO mode and experiences light traffic. Therefore, $\textit{WCET}^{\textit{LO,2}}$ is considered when scheduling tasks. Now, when the system is encountering heavy traffic, the execution-times of the object detection function increase (more than $\textit{WCET}^{\textit{LO}}$ of light traffic~(i.e., $\textit{WCET}^{\textit{LO,2}}$)). Therefore, an upper-level $\textit{WCET}^{\textit{LO}}$~(i.e., $\textit{WCET}^{\textit{LO,1}}$), set higher than the actual execution-times, is considered to handle these situations. In addition, if the actual execution times exceed $\textit{WCET}^{\textit{LO,1}}$, the system switches to the HI mode, and $\textit{WCET}^{\textit{HI}}$ is considered when scheduling tasks. 

% \textbf{QoS improvement in MulTi-MiCS approach:} 
To improve QoS in MulTi-MiCS, it is evident that 
lowering the $\textit{WCET}^{\textit{LO}}$ levels of HC tasks generates additional processor capacity, enabling scheduling more LC tasks for a period of time and improving QoS (like when the system encounters light traffic). 
When the $\textit{WCET}^{\textit{LO}}$ for a task is adjusted, the QoS of LC tasks~(for utilization bound guarantee) is also adjusted without affecting the $\textit{WCET}^{\textit{LO}}$ of HC tasks. 
This means that QoS can improve at run-time, depending on system behavior (compared to what is considered at design-time) and generated additional processor capacity. %but it will stay higher than the minimum acceptable level. 
Conversely, when the $\textit{WCET}^{\textit{LO}}$ levels of HC tasks increase (such as during a transition from light to heavy traffic), QoS of LC tasks decreases. Even if an LC task is currently being executed, it may be dropped to meet the real-time requirements of HC tasks. 
However, QoS will remain higher than the minimum acceptable level used in the HI mode. 
Note that the run-time adjustment timing overhead has been measured on the hardware platform ($\leq 1\mu S$) and considered in the proposed approach at design-time.

\vspace{-6pt}
\subsection{System Objectives Analysis}
\label{subsec:objanalys}
\vspace{-6pt}
System timing behavior is improved by pursuing two primary objectives:~\cite{Ranjbar21}: 
\begin{itemize}[leftmargin=*]
    \item \underline{Resource Utilization} is improved by achieving a substantial increase in utilization that can be assigned to LC tasks in the LO mode ($U_{LC}^{LO}$). If $U_l^m$ denotes total utilization of tasks with the criticality level~$l$~($l\in\{LC,HC\}$) in mode~$m$~($m\in\{LO,HI\}$), $U_l^m=\sum_{i\in\{LC,HC\}}{WCET^{k}_{i}/P_i}$. Employing the existing MC scheduling algorithm, EDF with Virtual Deadline~(EDF-VD)%~\cite{baruah12}, which has been used in numerous studies over the past decade, including
    ~\cite{baruah12,Liu2018,Gu2018,Ranjbar21}, 
    Eq.~(\ref{eq:utilobj}) shows $U_{LC}^{LO}$ which is upper-bounded by the schedulability constraints in both LO and HI modes, where all LC tasks are dropped in the HI mode~\cite{baruah12,Ranjbar21}. $U_{HC}^{LO}$ is computed based on $\textit{WCET}^{\textit{LO}}$ of HC tasks, in which $\textit{WCET}^{\textit{LO}}$s are computed according to \textit{AnTi-MiCS} approach~(i.e., Eq.~\eqref{eq:Metric1}-\eqref{eq:metricdriv}).  %In the following, it is assumed that all LC tasks are dropped in the HI mode.
    \begin{equation}
    \footnotesize
    \label{eq:utilobj}
        U_{LC}^{LO}\leq min\{1-U_{HC}^{LO},\frac{1-U_{HC}^{HI}}{U_{HC}^{LO} +\gamma(1-U_{HC}^{LO})}\}
    \end{equation}
    
    \item \underline{Mode Switching Probability} reduction has a beneficial effect on the performance or functionality of MC systems by reducing frequent drops of LC tasks in the HI mode. If $P^{ovrun}_{i}$ is the exceeding probability of task $\tau_i$ from the higher $WCET^{LO}_i$ level, the system mode switching probability ($P^{MS}_{Sys}$) is computed as follows~\cite{Ranjbar21}. 
    Let $P^{NMS}_{Sys}$ denote the probability that no HC task experiences an overrun, thereby preventing any mode switches. Consequently, the probability of a mode switch occurring is given by $P^{MS}_{Sys}= 1 - P^{NMS}_{Sys}$. Given that tasks are independent, $P^{MS}_{Sys}$ is determined using Eq.~(\ref{eq:PMSobj}).
    Note that $WCET^{LO}_i$ and $P^{ovrun}_{i}$ are computed according to \textit{AnTi-MiCS} approach. %, as discussed in Section~\ref{subsec:PPMethod1}.
    \begin{equation}
    \footnotesize
    \label{eq:PMSobj}
        P^{MS}_{Sys} = 1-\prod_{\zeta_i\in HC}(1-P^{ovrun}_{i})
    \end{equation}
\end{itemize}

To improve the system timing behavior by using these two objectives, Eq.~(\ref{eq:maxobj}) should be maximized~\cite{Ranjbar21}. %As a result, 
Analogous to~\cite{Ranjbar21,Ranjbar2021-tcad}, an approach capable of obtaining appropriate values for $\textit{WCET}^{\textit{LO}}$, should yield a higher value for this equation.
\begin{equation}
\footnotesize
    \label{eq:maxobj}
    max(U_{LC}^{LO}\times(1-P^{MS}_{Sys}))
\end{equation}

\vspace{-6pt}
\section{Experimental Evaluation}
\label{Sec:Evaluation}
\vspace{-6pt}
\vspace{-2pt}
\subsection{Experimental Setup}
\label{subsec:setup}
\vspace{-6pt}

Experiments were conducted on the ODROID-XU4 board, featuring ARM's big.LITTLE architecture with four Cortex-A15 (big) and four Cortex-A7 (LITTLE) cores. 
For our experiments, we used the LITTLE cores at a maximum frequency of 1.4 GHz. To evaluate our approach, we employed various benchmarks from the MiBench suite~\cite{Guthaus2001}, including those in the automotive, network, and telecommunication categories, such as \guilsinglleft insert-sort\guilsinglright, \guilsinglleft matrix-mult\guilsinglright, \guilsinglleft qsort\guilsinglright, \guilsinglleft bitcount\guilsinglright, \guilsinglleft dijkstra\guilsinglright, \guilsinglleft FFT\guilsinglright,
\guilsinglleft corner\guilsinglright, \guilsinglleft edge\guilsinglright, \guilsinglleft smooth\guilsinglright, and \guilsinglleft epic\guilsinglright, as well as the \guilsinglleft matrix-multiplier\guilsinglright~benchmark from AXBench~\cite{axbench}. Each benchmark was executed with various inputs on the board to determine their execution times. 

We evaluate the effectiveness of our approach by comparing it with state-of-the-art methods~\cite{baruah12,Ranjbar21,Ranjbar2023dac}. As detailed in Section~\ref{Sec:RelWork} and papers like~\cite{Ranjbar21}, most approaches, including~\cite{baruah12,Su2016,Liu2018,Gu2018}, drive ${WCET}^{LO}$, as a percentage of ${WCET}^{HI}$. We therefore select~\cite{baruah12} as a representative method and perform experiments using three different percentages of $\textit{WCET}^{\textit{HI}}$, as $\textit{WCET}^{\textit{LO}}$~($\lambda$=$\frac{\textit{WCET}^{\textit{LO}}}{\textit{WCET}^{\textit{HI}}}$).

\vspace{-4pt}
\subsection{MulTi-MiCS Evaluation}
\vspace{-6pt}

\begin{figure}[t]
	\centering
    \vspace{-9pt}
	\includegraphics[width=0.72\columnwidth]{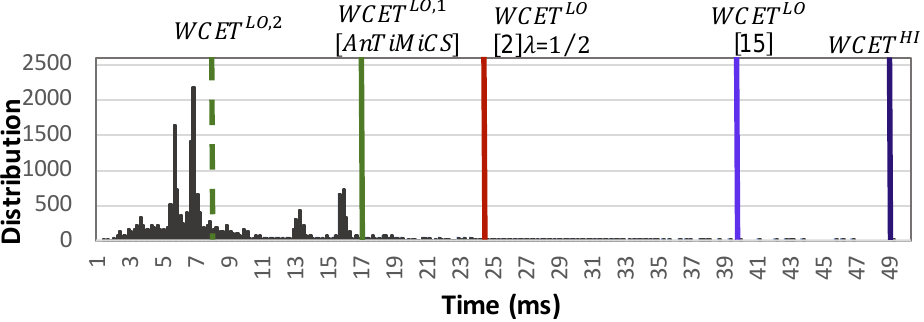}	
    \vspace{-9pt}
	\caption{MulTi-MiCS evaluation by analyzing Smooth application~\cite{Guthaus2001}.}
	\label{fig:ResultAnalyis}
 \vspace{-12pt}
\end{figure}

We first evaluate MulTi-MiCS in obtaining diverse $\textit{WCET}^{\textit{LO}}$ and its impact on improving utilization. In Fig.~\ref{fig:ResultAnalyis}, we depict the execution time distribution of the Smooth application~\cite{Guthaus2001}, characterized by more than one peak in its distribution curve. 
This application performs image smoothing and includes threshold, brightness, and spatial control parameters. Small input data may consist of black and white images (i.e., takes less time), while large input data can be more complex images (i.e., takes more time). 
Through using the proposed approach, AnTi-MiCS, we obtain $WCET^{LO,1}=16.6\text{ms}$ with a corresponding probability of overrun, $P^{ovrun}=9.02\%$. In the case that the task's execution time exceeds 16.6\textit{ms} and the output is not yet ready, the task experiences an overrun, leading to a transition to the HI mode.  \cite{Ranjbar21} and~\cite{baruah12} with $\lambda$=$\frac{1}{2}$ lead to processor under-utilization due to the selection of inappropriate $\textit{WCET}^{\textit{LO}}$. Using $\lambda$=$\frac{1}{8}$ in \cite{baruah12} elevates the likelihood of mode switches. Despite the proximity of the outcome to that of AnTi-MiCS when considering $\lambda$=$\frac{1}{4}$, it lacks a general formula applicable to other benchmarks to yield similar results. 
Next, considering the presence of multiple peaks and large variation in time distribution and assuming a temporal correlation between subsequent inputs presented to the applications, we employ the MulTi-MiCS approach to determine an additional level of $\textit{WCET}^{\textit{LO}}$, denoted as $WCET^{LO,2}=9.7\text{ms}$. Examining the application's distribution curve, we observe that 67.01\% of the time, the execution time of samples is less than 9.7\textit{ms} (e.g., in the scenario where only black and white images are provided as inputs), and when the input behavior changes, in 23.97\% of the times, $\textit{WCET}^{\textit{LO}}$ can equal to 16.6\textit{ms} (e.g., in the scenario where only complex images are provided as inputs). %It is important to note that the effectiveness of the MulTi-MiCS approach is particularly evident in scenarios characterized by temporal correlation between subsequent inputs presented to the application.
As a result, by considering two levels of $\textit{WCET}^{\textit{LO}}$, specifically when $\textit{WCET}^{\textit{LO}}=9.7ms$, the utilization can be improved by scheduling more LC tasks. Hence, the task utilization is defined by $\frac{Execution-Time}{Period}$~\cite{baruah12,burns2022mixed,Ranjbar21}. Based on the traditional model, if we consider one level of $\textit{WCET}^{\textit{LO}}$ and the task period 120\textit{ms}, the task utilization is $\frac{16.6}{120}=0.139$.
In contrast, in our proposed approach, incorporating two levels of $\textit{WCET}^{\textit{LO}}$, the task utilization can be $\frac{9.7}{120}=0.081$ for 67.01\% of the time. Consequently, idle time of $\frac{6.9}{120}=0.051$ is generated, which can be utilized to schedule and execute LC tasks. 
Analyzing all HC tasks leads to additional idle time generation, facilitating the scheduling of more LC tasks and thereby improving processor utilization and QoS. 
%It is important to note that, due to the proximity of the last and first two peaks, 
It is worth noting that, given the proximity of the last and first two peaks, 
designating a separate WCET level between them would be inefficient—a challenge effectively addressed by our approach.
%a consideration effectively addressed by our approach.

% Furthermore, the object detection function, explained in the motivational example of Section~\ref{Sec:Intro}, may encounter varying traffic loads during different hours. $WCET^{LO}$ can be set to 2.12\textit{s} when the automotive is experiencing heavy traffic. Conversely, it can be set to 0.38\textit{s} during light traffic periods, allowing for the scheduling of additional tasks in the generated idle time.

\vspace{-4pt}
\subsection{Evaluation with Real Benchmarks Model}
\vspace{-6pt}

\begin{table}[t]
    \centering
     \vspace{-14pt}
    \caption{Overrun percentage~($P^{ovrun}$) of benchmarks for different approaches}
    \footnotesize
    % \vspace{-10pt}
    \adjustbox{max width=\linewidth}{
    \begin{tabular}{|c|c|c|c|c|c|}%c|}
    \hline
    \footnotesize
         Benchmarks & MulTi-MiCS & \cite{Ranjbar21}, \cite{Ranjbar2023dac} & \cite{baruah12} $\lambda=\frac{1}{4}$ & \cite{baruah12} $\lambda=\frac{1}{8}$  & \cite{baruah12} $\lambda=\frac{1}{16}$\\% & Ranjbar'23  \\
        %  & & Ranjbar'23~\cite{Ranjbar2023dac} & \cite{baruah12} $\lambda=\frac{1}{4}$ & \cite{baruah12} $\lambda=\frac{1}{8}$ & \cite{baruah12} $\lambda=\frac{1}{16}$ \\%& \cite{Ranjbar2023dac}  \\
         \hline
         insertsort  & 3.87\% & 0.47\% & 0.01\% & 0.48\% & 95.93\%    \\
         quicksort  & 7.19\% & 0.33\% & 0.001\% & 0.02\% & 5.09\%    \\
         edge  & 2.29\% & 0.28\% & 0.35\% & 1.04\% & 2.75\%    \\
         epic  & 4.13\% & 0.43\% & 1.21\% & 22.31\% & 42.81\%    \\
         smooth  & 9.02\% & 0.003\% & 28.83\% & 60.05\% & 96.54\%    \\
         FFT  & 2.37\% & 0.28\% & 0.002\% & 0.01\% &0.01\%     \\
         matrixmult & 9.59\% & 0.02\% & 0.03\% & 0.05\% & 4.58\%    \\
         bitcount & 6.28\% & 0.24\% & 0.008\% & 0.02\% & 9.94\%   \\
         dijkstra & 4.15\% & 0.89\% & 0.07\% & 0.90\% & 100\%   \\
         \hline
    \end{tabular}
    }
    \label{tab:overrun}
    \vspace{-11pt}
\end{table}

We now present the effectiveness of \textit{AnTi-MiCS} in determining the appropriate $\textit{WCET}^{\textit{LO}}$ and the probability of overrunning it for different benchmarks. Table~\ref{tab:overrun} presents the percentage of task overrunning from $\textit{WCET}^{\textit{LO}}$ for each benchmark in~\cite{Guthaus2001,axbench}, mentioned in Section~\ref{subsec:setup}, under various policies. %, as outlined in related works. %, providing an assessment of different approaches. 
The table highlights that the approach of~\cite{baruah12} is inefficient for obtaining $\textit{WCET}^{\textit{LO}}$. Determining it as a percentage of $\textit{WCET}^{\textit{HI}}$ (e.g., $\lambda$=$\{\frac{1}{2},\frac{1}{4},\frac{1}{8},\frac{1}{16}\}$) proves impractical for establishing a general function for the probability of overrun since there are various overrunning probabilities for each $\lambda$ value. Moreover, the approach proposed by~\cite{Ranjbar21} determines $\textit{WCET}^{\textit{LO}}$ based on the applications' Average-Case Execution Time~(ACET) and the standard deviation~($\delta$) ($\textit{WCET}^{\textit{LO}}=ACET+n\delta$). Although a general formula is mentioned in~\cite{Ranjbar21} for the probability of overrun, the major drawback of~\cite{Ranjbar21} is 1)~the necessity to tune the parameter \textit{n} for each set of applications intended to run on a processor, requiring prior knowledge of which applications will execute on the same processor, 2) determining the mode switching probability is very pessimistic. In addition, \cite{Ranjbar2023dac} has followed the same policy as~\cite{Ranjbar21} for its design-time analysis. In contrast, our approach can be individually applied to any application for determining $\textit{WCET}^{\textit{LO}}$, which is also more realistic. 
However, to demonstrate the efficacy of our analytical approach in determining $\textit{WCET}^{\textit{LO}}$ compared to other methods, in the subsequent analysis, we analyze the objectives and system goal, outlined in Section~\ref{subsec:objanalys}.

Table~\ref{tab:metricresultoffline} outlines objectives for different approaches. 
%$\textit{P}^{\textit{MS}}_{\textit{Sys}}$ is the system mode switching probability~(i.e., Eq.~(\ref{eq:PMSobj})) if we set $\textit{WCET}^{\textit{LO}}$ according to different approaches. 
As discussed in Section~\ref{subsec:objanalys}, %$\textit{P}^{\textit{MS}}_{\textit{Sys}}$ and $max(U^{LC}_{LO})$ are the maximum utilization that can be assigned to LC tasks, and 
$max(U^{LO}_{LC})\times(1-P^{MS}_{Sys})$ is system goal. In this experiment, since we degrade LC tasks in the HI mode~(no dropping), $U^{LO}_{LC}$ is upper bounded by the formula mentioned in~\cite{Ranjbar2023dac,Liu2016}. 
These parameters are computed based on the design-time system analysis. For this experiment, \guilsinglleft insert-sort\guilsinglright, \guilsinglleft qsort\guilsinglright, \guilsinglleft smooth\guilsinglright, \guilsinglleft epic\guilsinglright, and \guilsinglleft edge\guilsinglright~are classified as HC tasks, and other benchmarks mentioned in Section~\ref{subsec:setup}, as LC tasks. 
We observe that although the mode switching probability in \cite{Ranjbar21,Ranjbar2023dac} and \cite{baruah12} with $\lambda$=$\frac{1}{2},\frac{1}{4}$ is less than AnTi-MiCS, their maximum utilization that can be assigned to LC tasks is worse in comparison with AnTi-MiCS, due to choosing $\textit{WCET}^{\textit{LO}}$ too pessimistic. Besides, although the maximum assigned utilization in AnTi-MiCS is almost equal to the scenario of \cite{baruah12} with $\lambda$=$\frac{1}{16}$, the mode switching probability is significantly better (i.e., lower) in the AnTi-MiCS scheme. In addition, both $P^{MS}_{Sys}$ and $max(U^{LC}_{LO})$ in \cite{baruah12} with $\lambda$=$\frac{1}{8}$ are worse than AnTi-MiCS, due to choosing $\textit{WCET}^{\textit{LO}}$ too optimistic. 
%As a result, AnTi-MiCS performs better than other approaches in making a reasonable trade-off between mode switching and utilization, and this fact is shown by having a higher value in system goal, compared to other approaches.
As a result, AnTi-MiCS outperforms other approaches by achieving a better trade-off between mode switching and utilization, as evidenced by its higher system goal value.

\begin{table}[t]
    \centering
    \begin{minipage}{0.48\textwidth}
        \centering
    \vspace{-18pt}
    \caption{System Performance at design-time for different approaches}
    \footnotesize
    \def\arraystretch{1.1}
    \adjustbox{max width=\linewidth}{
    \begin{threeparttable}
    \begin{tabular}{cccc}
    \hline
         % Metric & MulTi-MiCS & Ranjbar'21 & Barauh'12 & Barauh'12  & Barauh'12 & Ranjbar'23  \\

         & ${P^{MS}_{Sys}}^{**}$ & ${max(U^{LO}_{LC})}^{*}$ & $max(U^{LO}_{LC})\times$ \\
         &  & & $(1-P^{MS}_{Sys})$ \\
        \hline
        AnTi-MiCS & 23.96\% & 81.41\% & 0.619 \\

        \cite{Ranjbar21}, \cite{Ranjbar2023dac} & 1.50\% & 56.83\% &  0.560
        \\
        \cite{baruah12} $\lambda=\frac{1}{2}$ & 2.07\% & 50.71\% & 0.497  \\ 
        \cite{baruah12} $\lambda=\frac{1}{4}$ & 21.15\% & 63.32\% & 0.499 \\
        \cite{baruah12} $\lambda=\frac{1}{8}$ & 69.43\% & 73.47\%  & 0.225  \\ 
        \cite{baruah12} $\lambda=\frac{1}{16}$ & 100\% & 82.96\% &  0.000 \\
        \hline
       
         %  Metric & AnTi-MiCS & \cite{Ranjbar21}\cite{Ranjbar2023dac} & \cite{baruah12} $\lambda=\frac{1}{2}$ & \cite{baruah12} $\lambda=\frac{1}{4}$ &
         %  \cite{baruah12} $\lambda=\frac{1}{8}$& \cite{baruah12} $\lambda=\frac{1}{16}$   \\
         % \hline
         % $P^{MS}_{Sys}$ & 23.96\% & 1.50\% & 2.07\% & 21.15\% & 69.43\% & 100\%    \\
         % \hline
         % $U^{LO}_{LC}$ & 81.41\% & 56.83\% & 50.71\% & 63.32\% & 73.47\% &  82.96\%  \\
         % \hline
         % $U^{LO}_{LC}(1-P^{MS}_{Sys})$ & 0.619 & 0.560 & 0.497 & 0.499 & 0.225 & 0.000   \\
         % \hline
    \end{tabular}
    \begin{tablenotes}
			\item $^{*}$ {\footnotesize Higher is better}
			\quad\quad $^{**}$ {\footnotesize Lower is better}
	\end{tablenotes}
    \end{threeparttable}
    }
    \label{tab:metricresultoffline}
    \vspace{-11pt}
    \end{minipage}
    \hfill
    \begin{minipage}{0.48\textwidth}
        \centering
    \vspace{-10pt}
    \caption{System Performance at run-time for different approaches}
    \footnotesize
    \def\arraystretch{1.1}
    \adjustbox{max width=\linewidth}{
    \begin{threeparttable}
    \begin{tabular}{cccc}
    \hline
         % Metric & MulTi-MiCS & Ranjbar'21 & Barauh'12 & Barauh'12  & Barauh'12 & Ranjbar'23  \\
         & ${Util^{wst}_{Sys}}^{**}$ & $QoS^{*}$ & ${\#MS^{Sys}}^{**}$ \\
         \hline
         MulTi-MiCS & 30.42\% & 90.91\% & 28 \\ 
         AnTi-MiCS & 38.65\% & 84.50\% & 28 \\ 
         \cite{Ranjbar21} & 69.03\% & 62.69\% & 3\\ 
         \cite{Ranjbar2023dac} & 48.65\% & 71.66\% & 43 \\ 
         \cite{baruah12} $\lambda=\frac{1}{2}$ & 81.86\% & 56.52\% & 5 \\ 
         \cite{baruah12} $\lambda=\frac{1}{4}$ & 71.01\% & 57.77\% & 55 \\ 
         \cite{baruah12} $\lambda=\frac{1}{8}$ & 60.98\% & 54.55\% & 101\\ 
         \hline

    \end{tabular}
    \begin{tablenotes}
			\item $^{*}$ {\footnotesize Higher is better}
			\quad\quad $^{**}$ {\footnotesize Lower is better}
	\end{tablenotes}
 \end{threeparttable}
    }
    \label{tab:metricresultonline}
    \vspace{-8pt}
    \end{minipage}
    \vspace{-6pt}
\end{table}

Table~\ref{tab:metricresultonline} presents the run-time evaluation metrics for different approaches. \textit{QoS} denotes the percentage of executed LC task instances during run-time, and $MS^{Sys}$ indicates the average number of mode switches per hyper-period~\cite{Ranjbar2023dac}. $Util^{wst}_{Sys}$ represents the processor’s wasted utilization—i.e., the average percentage of the difference between tasks' WCET and their ACET, relative to their WCET. 
%presents the unused utilization on the processor~(i.e., wasted utilization or the average percentage of the difference between the WCET and actual execution times to WCET of tasks). %generated free slack on the processor that has not been used. 
Although $MS^{Sys}$ is higher in MulTi-MiCS and AnTi-MiCS compared to~\cite{Ranjbar21} and~\cite{baruah12} with $\lambda$=$\frac{1}{2}$, they achieve notably better performance by executing more LC tasks overall, resulting in higher QoS. This achievement is facilitated by determining appropriate WCET levels based on thorough distribution analysis of tasks, evident in the observed $Util^{wst}_{Sys}$ values compared to other approaches. 
Additionally, although \cite{baruah12} with $\lambda$=$\{\frac{1}{4},\frac{1}{8}\}$ assigns a high utilization to LC tasks at design-time, they exhibit less QoS due to more frequent mode switches. 
Besides, \cite{Ranjbar2023dac} performs better than other state-of-the-art works due to determining the proper $\textit{WCET}^{\textit{LO}}$s at run-time. However, both AnTi-MiCS and MulTi-MiCS approaches perform better than \cite{Ranjbar2023dac} due to 1)~well distribution analysis and diverse $\textit{WCET}^{\textit{LO}}$ levels determination, 2) no run-time timing overhead, \cite{Ranjbar2023dac} has a significant run-time timing overhead to adjust $\textit{WCET}^{\textit{LO}}$. 
MulTi-MiCS improves QoS by up to 36.36\%, and 30.27\% on average, compared to other approaches. By utilizing diverse $\textit{WCET}^{\textit{LO}}$ levels, it reduces wasted utilization by 8.23\% and enhances QoS by 6.41\% over AnTi-MiCS, which uses a single $\textit{WCET}^{\textit{LO}}$ level.
%Due to employing diverse $\textit{WCET}^{\textit{LO}}$ levels, MulTi-MiCS can reduce wasted utilization by 8.23\% and consequently improve QoS by 6.41\%, compared to AnTi-MiCS, which has one $\textit{WCET}^{\textit{LO}}$~level.

% \begin{table*}[t]
%     \centering
%     \caption{System Performance at run-time for different approaches}
%     % \vspace{-10pt}
%     \adjustbox{max width=\linewidth}{
%     \begin{threeparttable}
%     \begin{tabular}{|c|c|c|c|c|c|c|c|}
%     \hline
%          % Metric & MulTi-MiCS & Ranjbar'21 & Barauh'12 & Barauh'12  & Barauh'12 & Ranjbar'23  \\
%            & MulTi-MiCS & AnTi-MiCS & \cite{Ranjbar21} & \cite{Ranjbar2023dac}  & \cite{baruah12} $\lambda=\frac{1}{2}$ & \cite{baruah12} $\lambda=\frac{1}{4}$ & \cite{baruah12} $\lambda=\frac{1}{8}$  \\
%          \hline
%          ${Util^{wst}_{Sys}}^{**}$ & 30.42\% & 38.65\% & 69.03\% & 48.65\% & 81.86\% & 71.01\% & 60.98\%   \\
%          \hline
%          $QoS^{*}$ & 90.91\% & 84.50\% & 62.69\% & 71.66\% & 56.52\% & 57.77\% & 54.55\%   \\
%          \hline
%          ${\#MS^{Sys}}^{**}$ & 28 & 28 & 3 & 43 & 5 & 55 & 101   \\
%          \hline
%     \end{tabular}
%     \begin{tablenotes}
% 			\item $^{*}$ { Higher is better}
% 			\quad\quad\quad\quad $^{**}$ { Lower is better}
% 	\end{tablenotes}
%  \end{threeparttable}
%     }
%     \label{tab:metricresultonline}
%     % \vspace{-12pt}
% \end{table*}

\vspace{-3pt}
\subsection{Evaluating Scheduling Approaches}
\vspace{-4pt}

%In the following experiment, 
We now evaluate and compare the schedulable task sets (acceptance ratio) of our approach against state-of-the-art methods~\cite{baruah12,Liu2016}, both with and without our scheme, as well as the policy from~\cite{Ranjbar21} for determining $\textit{WCET}^{\textit{LO}}$.~The~EDF-VD scheduling algorithm is utilized in both~\cite{baruah12,Liu2016}. In the case of mode switches,~\cite{Liu2016} executes all LC tasks in HI mode by reducing their WCET to 50\%, while~\cite{baruah12} drops all LC tasks in HI mode. Note that our scheme for selecting the proper $\textit{WCET}^{\textit{LO}}$ for HC tasks is versatile and applicable to~any scheduling algorithm to optimize resource utilization and mode switching probability. 
%, where $U_{bound}=max(U^{LO}_{HC}+U^{LO}_{LC},U^{HI}_{HC})$.
Following the task generation method in~\cite{Ranjbar2021-tcad,baruah12,Guo2015}, tasks are incrementally added at random until the utilization bound~($U_{bound}=max(U^{LO}_{HC}+U^{LO}_{LC},U^{HI}_{HC})$) reaches a target threshold. We vary $\textit{U}_{\textit{bound}}$ from 0.05 to 1 in 0.05 increments. 
Inspired by real execution times in \cite{Ranjbar2021-tcad}, we define $\textit{WCET}^{\textit{HI}}$ values within the range of [52,1142]\textit{ms}. The task periods are then determined based on task utilization and the corresponding $\textit{WCET}^{\textit{HI}}$ values.
Then, 1000 task sets are generated for each $\textit{U}_{\textit{bound}}$.
In this experiment, we assume an equal probability that a task is HC or LC.

\begin{figure}[t]
\centering
\vspace{-6pt}
\subfloat[\footnotesize Improvement in~\cite{Liu2016}]{\label{AcceptRate14}{\includegraphics[width=0.492\columnwidth]{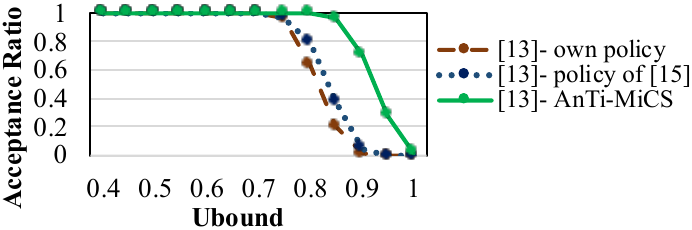}}}
% \hfill
\hspace{2pt}
\subfloat[\footnotesize Improvement in~\cite{baruah12}]{\label{AcceptRate1}{\includegraphics[width=0.492\columnwidth]{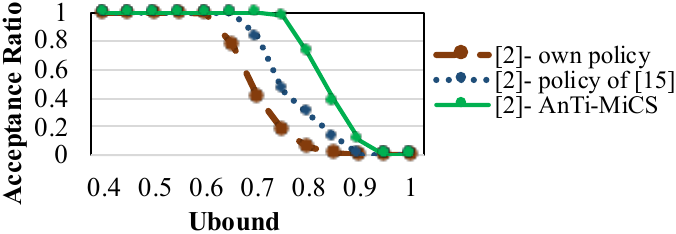}}}
\vspace{-8pt}
\caption{Acceptance ratio (schedulability) for two different scheduling approaches~\cite{baruah12}\cite{Liu2016} under two different policies of AnTi-MiCS and the policy proposed in~\cite{Ranjbar21}.}
\label{fig:AcceptRate}
\vspace{-10pt}
\end{figure}

Fig.~\ref{fig:AcceptRate} illustrates the acceptance ratio of two scheduling approaches from~\cite{baruah12,Liu2016} by varying utilization bounds consistent with prior works~\cite{baruah12,Liu2016,Ranjbar21}.
In Fig.~\ref{AcceptRate14}, for $U_{bound}$$\leq$$0.7$, all task sets are schedulable in~\cite{Liu2016} with three policies. As utilization increases (0.7<$U_{bound}$$\leq$0.8), AnTi-MiCS maintains full schedulability, whereas the acceptance ratio of \cite{Liu2016} with its own policy and \cite{Ranjbar21} policy decreases. Similarly, For $U_{bound}$>$0.8$, AnTi-MiCS outperforms the other two policies. This trend is consistent with~\cite{baruah12}, as depicted in Fig.~\ref{AcceptRate1}. 
%The reason for having a better acceptance ratio in AnTi-MiCS is the determination of the appropriate 
AnTi-MiCS achieves a higher acceptance ratio by accurately determining $\textit{WCET}^{\textit{LO}}$ for HC tasks, allowing more tasks to be scheduled. 
Consequently, it improves schedulability acceptance ratios of state-of-the-art MC algorithms by up to 68.76\%, and~42.19\% on average.
%As a result, AnTi-MiCS can enhance the schedulability acceptance ratio of state-of-the-art MC scheduling algorithms by up to~68.76\% and~42.19\% on average.

% \begin{figure}[t]
% 	\centering
% 	\includegraphics[width=0.4\columnwidth]{imgs/AcceptRate.pdf}	
%     \vspace{-10pt}
% 	\caption{Different scheduling approaches under two different policies; AnTi-MiCS and the policy proposed in~\cite{Ranjbar21}.}
% 	\label{fig:AcceptRate}
%  \vspace{-10pt}
% \end{figure}

\vspace{-9pt}
\section{Conclusion and Future Work}
\label{Sec:Conclude}
\vspace{-9pt}
% This article presented a framework for investigating task distributions and determining their WCET based on variation in task execution time distribution without incurring run-time timing overhead. The framework introduced a metric for obtaining a proper WCET, striking a reasonable tradeoff between utilization and mode switching, and then presented a scalable approach for obtaining diverse WCET levels to enhance utilization and QoS. %The proposed scheme improves the QoS by 30.27\% on average, coupled with a 35.89\% reduction in utilization waste, compared to other approaches.

This article proposed a framework for analyzing task execution time distributions and determining WCET without run-time overhead. It introduced a metric to balance utilization and mode switching, and presented a scalable method for deriving multiple WCET levels to improve utilization and QoS.

As future work, we would extend our framework by incorporating applications' run-time behavior and execution times to enhance utilization precision and QoS while maintaining mode switching probability. 
We would also extend the proposed approach to multiple criticality levels by establishing selection criteria for $\textit{WCET}^{\textit{l}}$ levels through the proposed MulTi-MiCS approach, where surpassing $\textit{WCET}^{\textit{l}}$ triggers a mode switch from mode \textit{l} to \textit{l}+1.

% While our approach targets dual-criticality systems to determine distinct $WCET^{LO}$ levels, it can be extended to multiple criticality levels. 
% This extension involves establishing selection criteria for $WCET^{l}$ levels through the proposed MulTi-MiCS approach, where surpassing $WCET^{l}$ triggers a mode switch from mode \textit{l} to \textit{l}+1. 
% As future work, we will analyze MC systems, scheduling policies, and timing overheads.
% %Analysis of MC systems, their scheduling policies, and timing overheads are mentioned as future work.
% %It should be note that, according to the input provided to the application, the time distribution of application can be updated at run-time, and the new values

\vspace{-7pt}

\subsubsection{\ackname} This work was supported in part by Deutsche Forschungsgemeinschaft~(DFG) through the Project \textit{LeanMICS} under Grant 534919862.

\bibliographystyle{splncs04}%{IEEEtran} %
\footnotesize
\vspace{-5pt}
\bibliography{Ref.bib}
% \vspace{-10pt}
% \begin{IEEEbiography}[{\includegraphics[width=0.92in,height=1.25in,clip]{imgs/Behnaz.JPG}}]{Behnaz Ranjbar} received the Ph.D. degree in computer science from Technische Universität~(TU) Dresden, Dresden, Germany in 2022. From October 2022 until March 2024, she was a Postdoctoral researcher at TU Dresden. She is currently a Postdoctoral researcher at the Chair of Embedded Systems, Ruhr University Bochum, Germany. Her research interests include real-time and low-power embedded system, fault-tolerant and reliable system design, and machine-learning-based embedded system design.
% \end{IEEEbiography}

% \begin{IEEEbiography}[{\includegraphics[width=0.92in,height=1.25in,clip]{imgs/akash.JPG}}]{Akash Kumar} (SM’13) received the joint Ph.D. degree in electrical engineering and embedded systems from the Eindhoven University of Technology, Eindhoven, Netherlands, and the National University of Singapore (NUS), Singapore, in 2009. From 2009 to 2015, he was with NUS. From October 2015 until March 2024, he was a Professor with Technische Universität Dresden, Germany, where he was directing the Chair of Processor Design. Since April 2024, he is directing the chair of Embedded Systems at Ruhr University Bochum, Germany. His research interests include the design and analysis of low-power embedded multiprocessor systems, and designing secure systems with emerging technologies.
% \end{IEEEbiography}

\end{document}